\documentstyle[aps,twocolumn]{revtex}
\begin{document}
\draft
\title{Wormholes supported by pure ghost radiation}
\author{Sean A. Hayward}
\address{Department of Science Education, Ewha Womans University, Seoul
120-750, Korea\\
\tt hayward@mm.ewha.ac.kr}
\date{17th February 2002}
\maketitle

\begin{abstract}
Traversible wormhole space-times are found as static, spherically symmetric
solutions to the Einstein equations with ingoing and outgoing pure ghost
radiation, i.e.\ pure radiation with negative energy density. Switching off the
radiation causes the wormhole to collapse to a Schwarzschild black hole.
\end{abstract}

The idea of space-time wormholes originated with Wheeler\cite{W}, with
traversible wormholes becoming popular after the article of Morris \&
Thorne\cite{MT}. More recently, renewed interest has focussed on outstanding
questions such as how to construct wormholes\cite{HKL}, their dynamical
behaviour\cite{wh}, their stability\cite{AP,BG} and the nature of the
negative-energy matter needed to support them\cite{DKMV}. This letter proposes
a very simple model of negative-energy radiation and exhibits resulting
wormhole solutions, which can be used to study the other questions.

The simplest radiation model is pure radiation, also known as null dust, for
which the energy tensor is $\rho u\otimes u$, where $u$ is a null vector and
$\rho$ is the energy density (which, if positive, can be absorbed in the
normalization of $u$). This represents incoherent radiation and occurs in the
geometric-optics limit of massless particles such as photons. The matter model
proposed here is simply pure radiation with negative energy density. In quantum
field theory, this is associated with ghost fields, so it may be called pure
ghost radiation. One might expect it to be a similarly valid model of
negative-energy radiation. More realistic models, like semi-classical quantum
fields such as those produced by Hawking radiation\cite{HPS}, are often not
analytically tractable, so a simple model has theoretical merit.

With pure radiation of the usual positive energy density, the static,
spherically symmetric Einstein system has been studied by Date\cite{D}, with
solutions found by Kramer\cite{K} and, more generally, by Gergely\cite{G}.
Reversing appropriate signs leads to the wormhole solutions derived below.
However, the geometry is quite different, in particular being non-singular. It
is curious that benign wormhole solutions naturally appear under such
situations.

Liberally adopting the method and notation of Gergely, the line element of a
static, spherically symmetric space-time may be written as
\begin{equation}\label{le}
ds^2=r^2d\Omega^2+f(r)^{-1}dr^2-h(r)dt^2
\end{equation}
where $r$ is the areal radius, $t$ is the time, $d\Omega^2$ refers to the unit
sphere and the metric functions $f$ and $h$ are positive. The active
gravitational mass-energy\cite{sph,1st} of such a space-time is
\begin{equation}
m(r)=(1-f(r))r/2.
\end{equation}
The energy tensor of pure ghost radiation is $-\tau u\otimes u$, where $u$ is a
null vector and $\tau\ge0$ is the negative energy density. With both ingoing
and outgoing radiation, the energy tensor is
\begin{equation}
T=-\tau_+u_+\otimes u_+-\tau_-u_-\otimes u_-
\end{equation}
where $u_\pm$ are null vectors. For the above metric, one may take
\begin{equation}
\sqrt2u_\pm={1\over{\sqrt{h}}}{\partial\over{\partial
t}}\pm\sqrt{f}{\partial\over{\partial r}}
\end{equation}
so that the null vectors are relatively normalized. Static solutions require
the same negative energy density $\tau=\tau_\pm$ for both ingoing and outgoing
radiation. This specialized matter model can also be regarded as an anisotropic
fluid with density $-\tau$, radial pressure $-\tau$ and vanishing tangential
pressure. Here and throughout, units are such that the speed of light and
Newton's constant are unity.

The energy-momentum conservation equations $\nabla\cdot T=0$ reduce to
\begin{equation}
(hr^2\tau)'=0
\end{equation}
where the prime denotes $d/dr$. Thus the (gravitationally red-shifted) radial
tension or negative linear mass density
\begin{equation}
\lambda=4\pi r^2h\tau
\end{equation}
is a positive constant. It is possible to absorb the magnitude of $\lambda$ in
the static Killing vector, but it will be retained here. In terms of the
function
\begin{equation}
\beta=-2\lambda/h
\end{equation}
the Einstein equations $G=8\pi T$ reduce to
\begin{eqnarray}
rf'&=&1-\beta-f\\
rfh'&=&h(1+\beta-f)\\
rf\beta'&=&-\beta(1+\beta-f)
\end{eqnarray}
where one equation is redundant, having already solved the conservation
equation. Transforming to $\rho=\ln r$ and using a dot to denote $d/d\rho$, the
reduced system is
\begin{eqnarray}
\dot f&=&1-\beta-f\\
f\dot\beta&=&-\beta(1+\beta-f).
\end{eqnarray}
The second equation yields
\begin{equation}
f={1+\beta\over{\beta-\dot\beta}}\beta
\end{equation}
which can be eliminated from the first equation to yield
\begin{equation}
(1+\beta)\beta\ddot\beta-(\beta+2)\dot\beta^2+\beta(1-2\beta)\dot\beta+2\beta^3=0.
\end{equation}
Dividing by $\beta^3$, this becomes an exact differential, integrating to
\begin{equation}
{1+\beta\over{\beta^2}}\dot\beta-{1\over\beta}-2\ln{\beta\over{r}}=D
\end{equation}
where $D$ is the integration constant. This leads to the algebraic relation
\begin{equation}
{(1+\beta)^2\over{2f\beta}}=\ln\left(-{r\over{a\beta}}\right)
\end{equation}
where $a=\exp(-(1+D)/2)$ is a positive constant. The branch of the logarithm
corresponds to negative $\beta$, reflecting the negative energy density.
Returning to the reduced system, one may eliminate $\rho$ or $r$ to give
\begin{equation}
{df\over{d\beta}}={f(\beta+f-1)\over{\beta(\beta-f+1)}}.
\end{equation}
Since solutions with negative $\beta$ are being sought, the method of Gergely
is now modified by defining
\begin{equation}
p=\sqrt{-2f\beta}\qquad l={1+\beta\over{\sqrt{-2f\beta}}}.
\end{equation}
Then the above equation becomes
\begin{equation}
{dp\over{dl}}=2(pl-1).
\end{equation}
Integrating the corresponding homogeneous equation yields
\begin{equation}
p=-2e^{l^2}\phi
\end{equation}
and the solution is completed by
\begin{equation}
\phi(l)=\int_0^le^{-\ell^2}d\ell+b={\sqrt\pi\over2}\hbox{erf}(l)+b
\end{equation}
where $b$ is constant and erf denotes the error function, the symmetric
integral of the standard normal distribution. The solution is mathematically
familiar compared with the case of positive energy density. Then
\begin{equation}
\beta=pl-1=-1-2le^{l^2}\phi.
\end{equation}
Collecting results, the class of solutions depends on the parameters $(a,b)$
and the normalizable $\lambda$, and may be summarized by
\begin{eqnarray}
r&=&-a\beta e^{-l^2}=a\left(e^{-l^2}+2l\phi\right)\\
h&=&-{2\lambda\over\beta}={2\lambda\over{1+2le^{l^2}\phi}}\\
f&=&-{p^2\over{2\beta}}={2e^{2l^2}\phi^2\over{1+2le^{l^2}\phi}}
\end{eqnarray}
as substituted into the line element (\ref{le}). By construction, this is the
unique class of static, spherically symmetric solutions to the Einstein
equations with pure ghost radiation.

It is straightforward to calculate
\begin{eqnarray}
{dr\over{dl}}&=&2a\phi\\
m&=&\left(e^{-l^2}+2l\phi-2e^{l^2}\phi^2\right)a/2\\
{dm\over{dl}}&=&-\left(1+2le^{l^2}\phi\right)a\phi.
\end{eqnarray}
The line element may then be written explicitly in $(t,l)$ coordinates as
\begin{eqnarray}
ds^2&=&a^2\left(e^{-l^2}+2l\phi\right)^2d\Omega^2
+2a^2e^{-l^2}\left(e^{-l^2}+2l\phi\right)dl^2\nonumber\\
&&-{2\lambda dt^2\over{1+2le^{l^2}\phi}}.
\end{eqnarray}
In the case $b=0$, $\phi(l)$ is an odd function and the metric is even in the
spatial coordinate $l$. This describes a symmetric wormhole with spatial
topology $R\times S^2$ and minimal surfaces at the wormhole throat $l=0$, with
radius $r=a$. The space-time is not asymptotically flat, but otherwise
constitutes a Morris-Thorne wormhole. There are no singularities, unlike in the
corresponding positive-energy solutions. The $b\not=0$ cases include asymmetric
wormholes which are analogous to the asymmetric Ellis wormholes\cite{E} for a
ghost Klein-Gordon field.

The solutions may be written in dual-null form
\begin{equation}\label{dn}
ds^2=r^2d\Omega^2-h\,dx^+dx^-
\end{equation}
in terms of null coordinates $x^\pm$ defined by
\begin{equation}
dx^\pm=dt\pm{a\over{\sqrt{\lambda}}}\left(e^{-l^2}+2l\phi\right)dl.
\end{equation}
Integration by parts yields an analytic solution:
\begin{equation}
x^\pm=t\pm{a\over{2\sqrt{\lambda}}}\left(le^{-l^2}+(1+2l^2)\phi\right).
\end{equation}
Thus the metric functions are implicitly known as functions of $x^+-x^-$. Using
the relations
\begin{equation}
\partial_\pm l=\pm{\sqrt{\lambda}\over{2a(e^{-l^2}+2l\phi)}}
\end{equation}
the solutions have been checked by substitution in the dual-null form of the
Einstein equations\cite{sph,1st} for the general spherically symmetric line
element (\ref{dn}) with non-zero energy components $T_{\pm\pm}=-h\tau/2$.

Returning to more general issues, the simplicity of pure ghost radiation means
that it is easier to study how the wormholes react to changes in the radiation
level. This has not been analytically tractable previously except in a
two-dimensional model\cite{HKL}. In particular, the ingoing and outgoing
radiation simply pass through one another without interaction, following null
geodesics with propagation equations $\partial_\pm(hr^2\tau_\pm)=0$. For
instance, it is not difficult to see what happens if the radiation supporting
the wormhole is turned off, as follows.

Consider a static wormhole for $\{x^+<0,x^-<0\}$, with the ghost radiation then
switched off from both sides of the wormhole, so that $\tau_\mp=0$ for
$x^\pm>0$, respectively. The solutions in the regions $\{x^\pm>0,x^\mp<0\}$
will be similar to Vaidya solutions\cite{V}, while the solution in the future
region $\{x^+>0,x^->0\}$ will be vacuum and therefore, by a version of
Birkhoff's theorem valid through the horizons of a black hole\cite{sph},
Schwarzschild. Further, this region can be seen to be be the interior of a
Schwarzschild black hole with mass $a/2$, by differentiability of the metric at
$x^+=x^-=0$. Specifically, continuity of the trapping horizons $\partial_\pm
r=0$\cite{sph,1st}, which initially form the throat $\{x^+=x^-,x^\pm<0\}$ of
the wormhole, means that they can be joined to the event horizons
$\{x^\pm=0,x^\mp>0\}$ of a black hole. This generally unexpected behaviour was
predicted by a general theory of wormholes and black holes\cite{wh} and
recently confirmed in the two-dimensional model\cite{HKL}. In summary, if the
supporting ghost radiation is switched off, the wormhole collapses to a
Schwarzschild black hole. Details of such collapses and other dynamical
processes involving the wormholes are to be presented in forthcoming work.

\medskip\noindent
Research supported by Korea Research Foundation grant KRF-2001-015-DP0095.
Thanks to Sung-Won Kim for support and Hisa-aki Shinkai for discussions.

\end{document}